\documentclass{aa}
\usepackage{graphics}
\usepackage{astrobib}
\usepackage{psfig}
\begin{document}

\thesaurus{03(11.03.1;11.03.4 Abell 1060;11.19.2;11.19.3)}

\title{H$\alpha$ Objective Prism Survey of Abell 1060}
\author{S.~M.~Bennett \inst{1}
	\and C.~Moss \inst{2}}
\institute{Institute of Astronomy, Madingley Road, Cambridge, CB3 0HA, UK \and  Blackett Laboratory, Imperial College of Science, Technology, and Medicine, Prince Consort Road, London, SW7 2BZ, UK}


\date{Received <date> / Accepted <date>}
\maketitle
\begin{abstract}

As part of a continuing study of the effect of cluster environment
on the star formation properties of galaxies, we have undertaken an
H$\alpha$ objective prism survey of the nearby cluster, Abell 1060.
We detect 33 galaxies in emission, 24 
of which are cluster members. We present new radial velocity measurements 
and  H$\alpha$+[\ion{N}{ii}] equivalent widths and fluxes for a number of these 
galaxies.  We distinguish between galaxies with
diffuse and compact emission, the latter having been associated in previous work with a disturbed morphology
of the galaxy and most likely resulting from tidally-induced star formation
from galaxy--galaxy or cluster--galaxy interactions.  The fraction of
cluster spirals in Abell 1060 detected with compact emission agrees with
the expected fraction for a cluster of its richness, as derived from results
of a previous survey of 8 clusters. Some of the detected cluster early-type
spirals exhibit anomalously high global H$\alpha$ equivalent widths, as compared
to galaxies of similar type in the field. 

\end{abstract}

\section{Introduction}
\label{resint}
\noindent

The effect of cluster environment on the star formation properties of 
galaxies has long been a matter of debate.
While some studies have suggested a lower star formation rate for cluster spirals
as compared to the field (e.g. \citeNP{gisler,dress85}), other work has suggested a similar or enhanced rate,
particularly for early-type spirals(e.g. \citeNP{kenn84,gav91,moss93,enacs}).
With the discovery that in distant rich clusters
there is a high fraction of blue, star-forming galaxies, often with unusual
morphology suggestive of mergers and tidal interactions (e.g. \citeNP{lavhen86,tom88,couch94}), there is renewed interest in
tidally-induced star formation by mergers and interactions in nearby clusters.
In order to address these issues, Moss, Whittle and co-authors have completed an 
objective-prism survey of eight nearby clusters of galaxies to detect global 
\mbox{H$\alpha$+[\ion{N}{ii}]} emission as an indicator of the current rate of massive 
star formation. The survey technique is described by \citeN{mwi}, hereafter MWI, 
and initial results have been discussed by \citeN{moss93} (see also 
\citeNP{moss90,moss95,moss96,moss97}). We have extended this survey to a ninth cluster, 
the Hydra I cluster, Abell 1060.

Abell 1060 is one of the nearest of the Abell clusters, at a redshift of $z\sim$0.01, 
and is the nearest large cluster beyond the Virgo and Fornax clusters. It has a high spiral 
fraction (e.g. \citeNP{solanes92}) and is a relatively poor cluster, with a low 
density intracluster medium \cite{lowmush} and low X-ray luminosity \cite{edge91}. 
Since it is the nearest of the clusters surveyed by us so far, it can be surveyed 
to a fainter limit in absolute magnitude. However, its proximity means that it 
has a large projected diameter on the sky, with one Abell radius,
\mbox{$1\, r_{A}=2\fdg3=1.5\, h^{-1}$ \rm{Mpc}} (where $h$ is defined in terms of 
the Hubble constant $H_{0}=100h\ \rm{km\ s}^{-1}\rm{Mpc}^{-1}$).
Whereas other clusters were surveyed in a region of radius 1.5 $r_{A}$, 
the photographic plate size restricted survey of Abell 1060 to a region of radius
somewhat less than one Abell radius (see \S \ref{platty}). 
\citeN{r89}, hereafter R89, presents a catalogue of 581 galaxies in the cluster 
area, which contains a sample which is complete to the magnitude limit 
$V_{25}=16.65$, within 2\degr \/ of the cluster centre.
 This is a convenient complete sample for the present H$\alpha$ survey,
extending to a fainter apparent magnitude than the Zwicky Catalogue used to 
define samples for other clusters.

Cluster properties are summarised in Table \ref{Hydratab}. The (B1950.0) position 
of the central cluster galaxy NGC 3311 is given as the cluster centre in columns 2 
and 3. The mean heliocentric radial velocity $\langle v_{\odot}\rangle$ and 
velocity dispersion $\sigma$ determined using $N_{gal}$ individual galaxy redshifts 
are given in columns 4, 5, and 6 \cite{bird94}. The Abell richness class \cite{aco} 
is given in column 7, the Bautz-Morgan and Rood-Shastry type classes are given in 
columns 8 and 9 respectively \cite{bm,strood}.

\begin{table*}

\centering

\caption{\label{Hydratab} Cluster properties}
\begin{tabular}{ll@{\hspace{1.0ex}}l@{\hspace{1.0ex}}ll@{\hspace{1.0ex}}l@{\hspace{1.0ex}}lcccccc}

\\
\hline
\\
Name & \multicolumn{6}{c}{R.A.~~(1950.0)~~Dec.} & $\langle v_{\odot}\rangle$ & \(\sigma\) & $N_{gal}$ &Richness&\multicolumn{2}{c}{Type Class} \\
\cline{12-13}
&\multicolumn{3}{c}{l}&\multicolumn{3}{c}{b}&km\,s$^{-1}$&km\,s$^{-1}$&&&B-M&R-S\\
\\

(1)&\multicolumn{3}{c}{(2)}&\multicolumn{3}{c}{(3)}&(4)&(5)&(6)&(7)&(8)&(9)\\
\\
\hline
\\
Abell 1060 &10$^{{\rm h}\hspace{-0.85ex}}$ &34$^{{\rm m}\hspace{-1.05ex}}$ &21\fs 6&-27$^{\circ\hspace{-0.85ex}}$ &16$^{\prime\hspace{-0.65ex}}$ &05$^{\prime\prime\hspace{-0.85ex}}$&3697 & 630 & 132 & 1 & III & C\\
(Hydra I)&\multicolumn{3}{c}{269\fdg6}&\multicolumn{3}{c}{26\fdg49}\\
\\
\hline
\\
\end{tabular}

\end{table*}

The observations and data reduction are described in \S \ref{obses}. Details 
of the observational technique are given in \S \ref{platty}, and of the process 
of identifying the emission-line galaxies in \S \ref{ident}, where a table of 
the detected emission-line galaxies (ELGs) is given. Measurements of radial velocities 
for the detected emission-line galaxies are presented in \S \ref{twoplate}, and 
those of \mbox{H$\alpha$+N[II]} equivalent widths and fluxes in \S \ref{ews}, where 
measured \mbox{H$\alpha$+N[II]} fluxes are also converted into effective 
star formation rates. 
A comparison of detected cluster emission-line galaxies in Abell 1060 with field galaxies
and detected emission-line galaxies in other clusters is given in 
\S \ref{frac}. Notes on individual galaxies are given in 
\S \ref{indiv}. Finally, we present a brief discussion of our results in 
\S \ref{discus}.

\section{\label{obses}Observations}

\subsection{\label{platty}Plate material}
\noindent Observations were made with the 61/94\,cm Curtis Schmidt at Cerro Tololo Inter-American Observatory equipped with a 6$^{\circ}$+4$^{\circ}$ objective prism combination. The plate scale of the Curtis Schmidt is \mbox{96.6 arcsec\,mm$^{-1}$}, with a usable field of 5$^{\circ}$ square. The dispersion at H$\alpha$ is \mbox{465$\pm$4\,\AA\,mm$^{-1}$}, as determined in \S \ref{twoplate}. At this moderate dispersion, the H$\alpha$ line is blended with the forbidden [\ion{N}{ii}] \mbox{($\lambda\lambda$=6548,6583\,\AA)} lines. Two exposures were made on hypersensitized IIIaF emulsion using an RG 630 filter. This combination of filter and emulsion gives an effective bandpass which has a FWHM of $\sim350$\,\AA ~centred on 6655\,\AA ~with a peak sensitivity at $\sim6717$\,\AA\,(MWI). At the relatively low redshift of the Hydra cluster, [\ion{S}{ii}] \mbox{($\lambda\lambda$=6716, 6731\AA)} emission can also be detected.

Repeated exposures of the same field taken with opposite dispersion directions are useful to confirm the reality of emission features allowing both elimination of spurious detections, and confirmation of weak emission features, particularly in cases where the underlying continuum is too faint to be seen. Reversing the dispersion direction also enables redshift determination (see MWI).
The seeing is critical to the detection of emission. Details of the exposures, both taken under good seeing conditions, are given in Table \ref{pltab}.

\begin{figure*}
\psfig{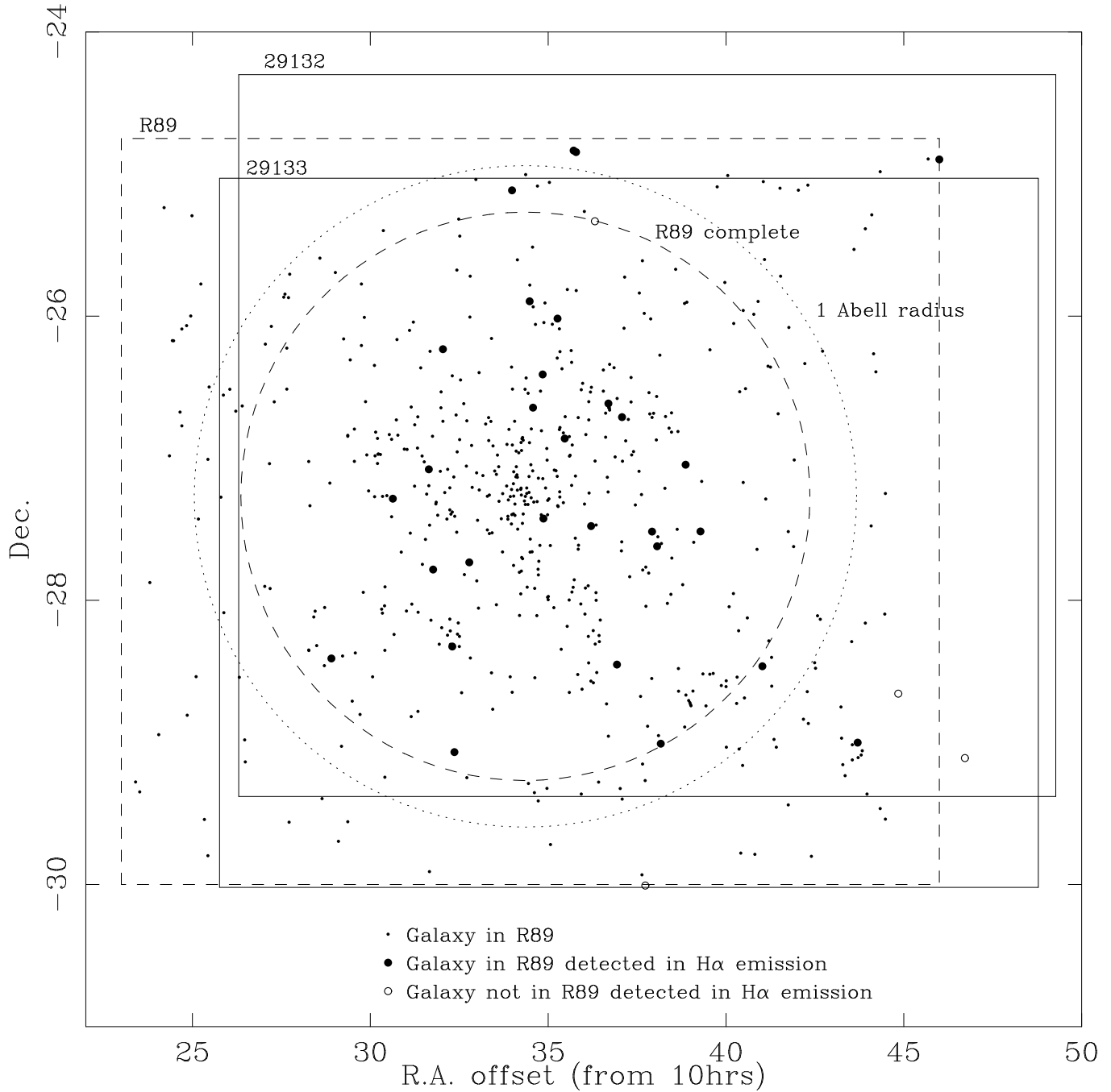}
\caption{\label{plfig} Survey coverage. Plate boundaries are indicated by solid lines. The area covered by R89 is shown by the dashed box, the area within which it is complete by the dashed circle. The dotted circle has a radius of 1.5 $h^{-1}$ Mpc (1 Abell radius) and is centred on the central cluster galaxy NGC 3311. Galaxies in R89 are shown as dots, detected emission-line galaxies as filled and open circles. } 
\end{figure*}

The fields covered by the two plates are not exactly coincident. The plate boundaries are shown in Fig. \ref{plfig} as solid boxes. The coverage of R89 is also shown: the dashed box shows the area covered by the catalogue, the dashed circle the area within which it is complete to the magnitude limit \mbox{$V_{25} = 16.65$}. One Abell radius is indicated by the dotted circle. There are three distinct regions in Fig. \ref{plfig}: the region in which the plates overlap, the region covered by only one plate, and the region within which R89 is complete. The complete region is contained entirely within the overlap region.
 
\begin{table*}
\centering
\caption{\label{pltab} Objective prism plates}
\begin{tabular}{ll@{\hspace{1.0ex}}l@{\hspace{1.0ex}}ll@{\hspace{1.0ex}}llccccc}
\\
\hline
\\
Plate& \multicolumn{3}{c}{U.T. date} & \multicolumn{3}{c}{Plate centre} & Filter & Emulsion & Exp. & Prism & Apex \\
number&&&& \multicolumn{3}{c}{R.A.~(1950.0)~Dec.}& & &  min.&&N/S\\
\\
(1)&\multicolumn{3}{c}{(2)}&\multicolumn{2}{c}{(3)}&(4)&(5)&(6)&(7)&(8)&(9)\\
\\
\hline
\\
29132&Feb&23&1985&10$^{{\rm h}\hspace{-0.85ex}}$  &	34$^{{\rm m}\hspace{-1.05ex}}$ &-27\fdg 5&RG 630&IIIaF&100& 6$^{\circ\hspace{-0.85ex}}$+4$^{\circ\hspace{-0.85ex}}$&N\\
29133&Feb&23&1985&10  &	34.5 &-27.3&RG 630&IIIaF&100&6+4&S\\
\\
\hline
\end{tabular}
\end{table*}

\subsection{\label{ident}Identification of emission-line galaxies}

The plates were searched systematically for galaxies showing H$\alpha$ emission using a low-power binocular microscope. A visibility parameter (S strong, MS medium-strong, M medium, MW medium-weak, or W weak) describing how readily the emission is seen on the plate, and a concentration parameter (VD very diffuse, D diffuse, N normal, C concentrated, or VC very concentrated) describing the spatial distribution of the emission and the contrast with the underlying continuum, were assigned to each candidate following the scheme of MWI. Within the overlap region galaxies were only accepted as emission-line sources if they had been independently detected on both plates and the separate emission features were consistent with each other. Outside the overlap area, galaxies were accepted only if they showed strong emission. Galaxies detected in H$\alpha$ are listed in Table \ref{elgtab} in order of increasing Right Ascension. 

In order to construct a complete sample of surveyed galaxies for statistical analysis, it is necessary to correct for the effect of overlapping spectra.  This is a particular problem at the relatively low galactic latitude \mbox{($b = 26\fdg5$)} of Abell 1060. All galaxies within the region of completeness of R89 brighter than the limiting magnitude $V_{25}=16.65$ were checked on both plates to ensure that their spectra were not overlapped by those of nearby objects. Table \ref{sur} lists both the identifications in R89 of the galaxies included in the complete sample, and those of galaxies which are deemed not to have been surveyed due to overlapping spectra. Those in the former list can be regarded as being confirmed as non emission-line galaxies if they do not appear in the table of detected emission-line galaxies (Table \ref{elgtab}).  

The objective prism spectra of the detected emission-line galaxies were digitised using the PDS (Plate Density Sensitometer) facility at the Royal Greenwich Observatory in order to measure line redshifts, equivalent widths, and fluxes.

\begin{table*}[!t]
\centering
\caption{\label{elgtab} Galaxies with detected H$\alpha$ emission}
\begin{tabular}{rrrrl@{\hspace{1.0ex}}l@{\hspace{1.0ex}}l@{\hspace{2.0ex}}l@{\hspace{1.0ex}}l@{\hspace{-1.0ex}}lclrrrc}
\\
\hline
\\
No. & \multicolumn{3}{c}{Catalogue name} & \multicolumn{6}{c}{R.A.~~(1950.0)~~Dec.~~~~} & $V$ & Type & \multicolumn{2}{c}{H$\alpha$ emission} & \multicolumn{1}{c}{$v_{\odot}$} & Notes \\
\cline{2-4} \cline{13-14}
 & R89 & ESO & IRAS & & & & & & & &R89 & Vis. & Conc. & \multicolumn{1}{c}{km\,s$^{-1}$} &  \\
 & W85 & NGC & Arp& & & & & & & &Wang & & & &  \\
\\
(1) & (2) & (3) & (4) & \multicolumn{3}{c}{(5)} & \multicolumn{3}{c}{(6)} & (7) & (8) & (9) & (10) & (11) & (12) \\
\\

\hline
\\
1 &	57 &	 	&10288-2824 &	10$^{ h\hspace{-0.85ex}}$ &	28$^{m\hspace{-1.05ex}}$ &	54\fs 2&	-28$^{\circ\hspace{-0.85ex}}$ &	24$^{\prime\hspace{-0.65ex}}$ &	31$^{\prime\prime\hspace{-0.85ex}}$ & 15.33 	 &\dots	 &MW	&N  &3689  	 &	  	\\
&&&&&&&&&&&SABa\\										           			        
2 &	101 &		 &	 &	10 &	30 &	38 &	-27 &	17.1 &	 &  15.86 	 &E	 &W	 &N &4800$^a$\hspace{-1.3ex}  	 &	  	\\
\\										           			        
3 &	128 &	501-G16 &10316-2704 &	10 &	31 &	38.7 &	-27 &	04 &	40 & 14.14 	 &SBab(rs)&MW	 &C &9683  	 &	  	\\
& & &&&&&&&&&SBa &\\										           			        
4 &	135 &		 &	 &	10 &	31 &	46 &	-27 &	47 &	01 & 15.39 	 &SB pec &W	 &N &3398  	 &	  	\\
\\										           			        
5 &	144 &	501-G17 &10320-2613 &	10 &	32 &	2.4 &	-26 &	13 &	56 & 14.18 	 &S0(7) &MS	 &C &4429  	 &	  	\\
& & &&&&&&&&&Sa(sp)\\										           			        
6 &	158 &	436-IG42 &10323-2819 &	10 &	32 &	18.3 &	-28 &	19 &	28 & 14.80 	 &E4:	 &S	&VC &3447  	 &*	  	\\
 &	68&&&&&&&&&&IG\\						           			         
7 &	161 &	436-G43 &	 &	10 &	32 &	21.9 &	-29 &	04 &	03 & 15.45 	 &S\dots &MW	&N &9123  	 &*	  	\\
\\										           			         
8 &	181 &	437-IG3 &	 &	10 &	32 &	46.9 &	-27 &	43 &	58 & 15.13 	 &S pec	 &MW	&D &2391  	 &*	  	\\
\\										           			         
9 &	229 &	501-G32 &	 &	10 &	33 &	59 &	-25 &	06.9 &	 &  14.26 	 &SBb(r) pec &MS&C &3933  	 &*	  	\\
\\										           			         
10 &	281 &		&	 &	10 &	34 &	28.8 &	-25 &	53 &	42 & 15.17 	 &Sa	 &M	&N &3686  	 &	  	\\
\\										           			         
11 &	291 &		 &	 &	10 &	34 &	34.5 &	-26 &	38 &	36 & 15.96 	 &S pec	 &M	&C &3170  	 &	  	\\
 &	70\\									           			         
12 &	310 &	501-G45 &10348-2624 &	10 &	34 &	50.8 &	-26 &	24 &	38 & 14.41 	 &S0(3)/a(r) &M	&N &4578  	 &*	  	\\
 &	71&&&&&&&&&&SAB(s)a\\									           			         
13 &	312 &	501-IG46 &10348-2725 &	10 &	34 &	51.9 &	-27 &	25 &	27 & 13.11 	 &S(B)c	 &S	&C &2851  	 &*	  	\\
 &	 &	3314A&1034-272&&&&&&&&(IG)\\							           			         
14 &	335 &	501-G53 &10352-2600 &	10 &	35 &	15.9 &	-26 &	01 &	00 & 14.26 	 &Sa:	 &MS	&D &3814  	 &	  	\\
& & &&&&&&&&&Sa\\										           			         
15 &	348 &	501-G59 &10354-2651 &	10 &	35 &	28 &	-26 &	51 &	40 & 13.55 	 &Sa	 &MS	&N &2385  	 &	  	\\
& & &&&&&&&&&SAa\\										           			         
16*\hspace{-1.2ex} &	363 &	501-IG61 &	 &	10 &	35 &	43 &	-24 &	50.1 &	 &  \dots 	 &``Multiple'' &S	&N &3597  	 &*	  	\\
\\										           			         
17*\hspace{-1.2ex} &	367 &	501-G62 &	 &	10 &	35 &	47 &	-24 &	50.7 &	 &  \dots 	 &Sb	 &S	&N &10200$^a$\hspace{-1.3ex}  	 &*	  	\\
 &	76\\									           			         
       
\\
\hline
\\
\end{tabular}
\end{table*}

\addtocounter{table}{-1}

\begin{table*}[!t]

\caption{\label{elgtableb} {Continued}}
\begin{tabular}{rrrrl@{\hspace{1.0ex}}l@{\hspace{1.0ex}}l@{\hspace{2.0ex}}l@{\hspace{1.0ex}}l@{\hspace{-1.0ex}}lclrrrc}

\\
\hline
\\
No. & \multicolumn{3}{c}{Catalogue name} & \multicolumn{6}{c}{R.A.~~(1950.0)~~Dec.~~~~} & $V$ & Type & \multicolumn{2}{c}{H$\alpha$ emission} & $v_{\odot}$ & Notes \\
\cline{2-4} \cline{13-14}
 & R89 & ESO & IRAS & & & & & & & &R89 & Vis. & Conc. & km\,s$^{-1}$ &  \\
 & W85 & NGC & Arp& & & & & & & &Wang  & & & &  \\
\\
(1) & (2) & (3) & (4) & \multicolumn{3}{c}{(5)} & \multicolumn{3}{c}{(6)} & (7) & (8) & (9) & (10) & (11) & (12) \\
\\
\hline
\\

18 &	390 &	501-G65 &10361-2728 &	10$^{ h\hspace{-0.85ex}}$ &	36$^{m\hspace{-1.05ex}}$ &	12\fs 3 &	-27$^{\circ\hspace{-0.85ex}}$ &	28$^{\prime\hspace{-0.65ex}}$&	37$^{\prime\prime\hspace{-0.85ex}}$ & 13.37 	 &Sbc	 &MS	&D &4378  	 &*	  	\\
 &	77&&1036-272&&&&&&&&Irr\\									           			         
19 &	&		 &	 &	10 &	36 &	18.9 &	-25 &	19 &	55 & \dots 	 &\dots	 &MW	&C &3000$^a$\hspace{-1.3ex} 	 &	  	\\
 &	78\\									           			         
20 &	413 &	501-G67 &10366-2636 &	10 &	36 &	41.7 &	-26 &	36 &	54 & 15.05 	 &SBa(s) &MW	&N &10594$^b$\hspace{-1.1ex} 	 &	  	\\
 &	79&&&&&&&&&&SBa\\									           			         
21 &	420 &		 &10369-2827 &	10 &	36 &	56.1 &	-28 &	27 &	08 & 15.83 	 &S0(3) &W	&N &2811  	 &	  	\\
&&&&&&&&&&&S0/a\\								
22 &	425 &	501-G70 &10370-2642 &	10 &	37 &	4.5 &	-26 &	42 &	40 & 14.99 	 &SBa(rs) &M	&C &10707 	 &	  	\\
&&&&&&&&&&&SB(r)0/a\\							

23*\hspace{-1.2ex} &	 &	437-G35 &10377-3000 &	10 &	37 &	44 &	-30 &	00 &	24 &\dots 	 &SB(s)bc: &S	&VC &3396$^c$\hspace{-1.0ex}  	 &*	  	\\
&&&&&&&&&&&SABa\\				           			         
24 &	457 &	437-G36 &10379-2730 &	10 &	37 &	55.4 &	-27 &	30 &	58 & 12.19 	 &S(B)c(rs) &MW	&D &3940  	 &*	  	\\
 &	 &	3336&1037-273&&&&&&&&SAc\\

25 &	461 &		 &F10381-2737 &	10  &	38 &	3.9 &	-27 &	37 &	12 & 14.26 	 &S0(2)	 &MW	&N &4503  	 &*	  	\\
&&&1038-273\\       			         
26 & 	465 & 	437-IG37&	&	10 &	38 &	10 &	-29 &	00.5 &	 & 15.98 &	``Double'' &M	&N &3600$^a$\hspace{-1.3ex}	&\\
\\

27 &	483 &		 &F10388-2702 &	10 &	38 &	52 &	-27 &	02.7 &	 &  16.05 	 &S	 &M	&N &11500$^a$\hspace{-1.3ex} 	 &	  	\\
\\										           			         
28 &	492 &	437-G43 &	 &	10 &	39 &	17 &	-27 &	30.9 &	 &  16.43 	 &SB\dots &M	&N &4100$^a$\hspace{-1.3ex}	 &	  	\\
\\										           			         
29 &	525 &		&	 &	10 &	41 &	1.4 &	-28 &	27 &	49 & 16.53 	 &S	 &W	&N &11200$^a$\hspace{-1.3ex} 	 &	  	\\
 &	82\\									           			         
30 &	564 &		 &	 &	10 &	43 &	42 &	-29 &	00.1 &	 &  15.23 	 &S	 &MW	&D &9600$^a$\hspace{-1.3ex} 	 &	  	\\
&&&1043-285\\										           			         
31 &	 &		 &10448-2839 &	10 &	44 &	50.7 &	-28 &	39 &	23 & 15.5 	 &\dots	 &S	&VC &10235$^d$\hspace{-1.0ex}  	 &	  	\\
&&&&&&&&&&&SBb\\										           			         
32*\hspace{-1.2ex} &	581 &	501-G100 &10459-2453 &	10 &	46 &	0 &	-24 &	53.8 &	 &  \dots 	 &Sa	 &S	&C &3720  	 &*	  	\\
 &	84 &	3393&1045-245&&&&&&&&SAab\\							           			         
33 &	 &	437-G64 &F10466-2906 &	10 &	46 &	43 &	-29 &	06 &	36 & \dots 	 &Sb$^2$	 &W	&N &4000$^a$\hspace{-1.3ex}  	 &	  	\\  
\\
\hline
\\      		    							         
\end{tabular}      		 					
\end{table*}
\addtocounter{table}{-1}

\begin{table*}[!t]
\caption{\label{elgtablec} {Notes to the Table}}


\noindent Column 1. Identification number. An asterisk indicates that the object lies outside the overlap region and has therefore only been detected on one plate.
         
Column 2. Line 1, identification in R89. Line 2, identification in \citeN{wpv}.
         
Column 3. Line 1, identification in \citeN{lau}. Line 2, NGC number.

Column 4. Line 1, IRAS identification from \citeN{wang} or NED. Line 2, identification in \citeN{arps}.
         
Columns 5 \& 6. (B 1950.0) Right Ascension and Declination. These are taken from R89 except no. 19 \cite{wpv}, no. 23 \cite{lau}, no. 31 \cite{vanvan}, no. 33 \cite{lau} and are quoted to the same precision as the source.
         
Column 7. $V$ magnitude, $V_{25}$ from R89.
         
Column 8. Morphological types from R89 (line 1) and \citeN{wang} (line 2) except no. 33 \cite{lau}. Wang's types should be more reliable than Richter's as they are based on a UK Schmidt IIIaJ plate. Note, though, that Wang's identification of the galaxy pair NGC 3314 as IG (interacting galaxy) is misleading (see text).
         
Column 9. Visibility parameter (S strong, MS medium-strong, M medium, MW medium-weak, W weak).

Column 10. Concentration parameter (VD very diffuse, D diffuse, N normal, C concentrated, VC very concentrated). 
         
Column 11. Heliocentric velocity. From R89, otherwise superscript indicates a)\ this paper, 
b)\ \citeN{rc3},
c)\ \citeN{mfb}, 
d)\ \citeN{strauss}.
         
Column 12. An asterisk indicates a note: {\em No. 7} appears to be tidally distorted. \citeN{lau} notes ``bright centre, faint disturbed envelope.''
{\em No. 8} has an apparently double nucleus, and could be a merger remnant.
{\em No. 9} has peculiar morphology, exhibits several distinct emission regions.
{\em No. 12} is asymmetric. \citeN{lau} notes an ``extremely faint ring.''
{\em No. 18}, noted by \citeN{arps} as a disturbed spiral, has a peculiar morphology and may have been tidally disrupted. Several emission regions are visible.
{\em No. 23} is noted as disturbed by \citeN{lau} and ascribed ``thick arms, knots, and dust patches, optical pair'' by \citeN{sgc}.
{\em No. 24}, NGC 3336, is a face-on Sc galaxy, with many individual HII regions visible in emission.  
{\em No. 25} is noted by Arp \& Madore as ``close pair, end of chain.''
{\em No. 32}, NGC 3393, is a type 2 Seyfert \cite{storch}. 
{\em Nos. 6, 13, 16, \& 17} see \S \ref{indiv}.
\newpage
\end{table*}

\begin{table*}
\centering
\caption{\label{sur} Identifications in R89 of galaxies included in the complete sample}
\begin{tabular}{lllllllllllllllllll}
\\
\hline
\\
\multicolumn{19}{l}{Galaxies surveyed}\\
\\
\hline
\\
  32 &  33 &  34 &  35 &  36 &  37 &  45 &  46 &  47 &  48 &  49 &  50 &  51 &  54 &  55 &  56 &  57 &  58 &  61 \\  
  62 &  64 &  65 &  66 &  67 &  69 &  70 &  71 &  72 &  73 &  74 &  75 &  76 &  77 &  78 &  79 &  80 &  81 &  82 \\
  83 &  85 &  86 &  88 &  90 &  91 &  94 &  97 &  98 &  99 & 100 & 101 & 103 & 104 & 105 & 106 & 109 & 110 & 112 \\
 113 & 114 & 115 & 116 & 117 & 119 & 120 & 121 & 122 & 124 & 126 & 128 & 129 & 131 & 132 & 133 & 134 & 135 & 138 \\
 139 & 144 & 145 & 148 & 149 & 150 & 154 & 155 & 156 & 158 & 161 & 162 & 164 & 165 & 166 & 167 & 168 & 169 & 171 \\
 174 & 175 & 177 & 179 & 181 & 182 & 185 & 186 & 188 & 189 & 190 & 191 & 192 & 193 & 194 & 196 & 198 & 202 & 203 \\
 204 & 206 & 208 & 209 & 210 & 211 & 212 & 213 & 214 & 215 & 216 & 217 & 218 & 219 & 220 & 221 & 222 & 224 & 225 \\
 226 & 227 & 232 & 233 & 234 & 236 & 237 & 238 & 239 & 241 & 243 & 244 & 245 & 246 & 247 & 249 & 252 & 253 & 254 \\
 255 & 256 & 258 & 260 & 261 & 264 & 266 & 268 & 269 & 271 & 273 & 274 & 278 & 279 & 281 & 282 & 283 & 286 & 288 \\
 289 & 290 & 291 & 292 & 293 & 295 & 296 & 297 & 299 & 303 & 305 & 306 & 307 & 308 & 310 & 312 & 313 & 314 & 316 \\
 317 & 319 & 320 & 322 & 325 & 327 & 329 & 333 & 334 & 335 & 336 & 338 & 339 & 340 & 341 & 342 & 343 & 344 & 345 \\
 347 & 348 & 351 & 353 & 357 & 358 & 359 & 360 & 365 & 368 & 369 & 373 & 375 & 376 & 377 & 379 & 381 & 384 & 385 \\
 386 & 387 & 389 & 390 & 391 & 396 & 397 & 400 & 403 & 405 & 407 & 409 & 413 & 416 & 418 & 419 & 420 & 421 & 423 \\
 425 & 427 & 428 & 431 & 433 & 437 & 439 & 440 & 443 & 445 & 446 & 450 & 453 & 457 & 461 & 463 & 464 & 465 & 466 \\
 469 & 471 & 474 & 475 & 477 & 478 & 479 & 480 & 481 & 482 & 483 & 484 & 485 & 486 & 487 & 489 & 490 & 491 & 492 \\
 493 & 494 & 495 & 496 & 497 & 498 & 500 & 502 & 506 & 507 & 509 & 511 & 513 & 516 & 517 & 519 & 520 & 521 & 523 \\
 525 & 528 & 529 & 530 & 531 & 532 & 540 & 546\\ 
\\

\hline
\\
\multicolumn{19}{l}{Galaxies not surveyed due to overlapping stellar spectra}\\
\\
\hline
\\

 40 &  41 &  84 & 123 & 141 & 159 & 170 & 173 & 183 & 205 & 223 & 228 & 230 & 231 & 250 & 287 & 311 & 330 & 337 \\ 
371 & 372 & 383 & 394 & 412 & 429 & 455 & 468 & 501 & 503 & 508 & 510 & 536 & 542 \\
\\
\hline
\end{tabular}
\end{table*}

\subsection{\label{twoplate}Redshift determinations}

As demonstrated by MWI, after \citeN{sto}, it is possible to obtain redshift estimates for emission-line galaxies accurate to within a few hundred km\,s$^{-1}$ using two Schmidt telescope objective prism plates taken with opposite dispersion directions. Following the method detailed in MWI, we obtained redshift estimates for all detected emission-line galaxies which we calibrated against available redshifts from the literature. Assuming that the dispersion of the prism combination is approximately constant over the narrow wavelength range of interest, the slope of this calibration gives a value for the dispersion of the  6$^{\circ}$+4$^{\circ}$ prism combination of \mbox{465$\pm$4 \hspace{0.15ex}\AA\hspace{0.35ex}mm$^{-1}$} between 6600 and \mbox{6800\,\AA}. The rms scatter about this relation of \mbox{$\sim$340 km\,s$^{-1}$} is slightly larger than that found by MWI of \mbox{$\sim$205 km\,s$^{-1}$}; this may be partly due to the fact that MWI used a two-dimensional cross-correlation technique, whereas we used the one-dimensional analogue.

Heliocentric velocities determined using the two-plate technique for seven galaxies which do not have existing redshift determinations and the emission associated with the superposed galaxies NGC 3314 (see \S \ref{indiv}) are listed in Table \ref{ztab}. 

\begin{table}
\centering
\caption{\label{ztab} Redshifts of emission-line galaxies}
\begin{tabular}{rrr}
\\
\hline
\\
No. & Galaxy name&\multicolumn{1}{c}{$v_{\odot}$}\\
& &\multicolumn{1}{c}{km  s$^{-1}$}\\
\\
(1)&(2)&(3)\\
\\
\hline
\\
2&R89 101&4800\\
13&NGC 3314A&2600\\
19&WPV 78&3000\\
27&R89 483&11500\\
28&ESO 437-G43&4100\\
29&R89 525&11200\\
30&AM 1043-285&9600\\
33&ESO 437-G64&4000\\
\\
\hline
\end{tabular}
\\
\end{table}

\subsection{\label{ews}Equivalent widths and fluxes}

\noindent
The two-dimensional digitised spectral scans were reduced to one-dimensional
spectra. These were used to determine equivalent widths of the blended 
\mbox{H$\alpha$+[\ion{N}{ii}]} emission lines 
using the dispersion value derived from the 
two-plate redshift determinations (\S \ref{twoplate} above).

It was only possible to measure a reliable equivalent width for 13 of the 33 
emission-line galaxies. Other galaxies had spectra which were overlapped by those of other objects, or emission or continuum which was too faint, or too diffuse, to be measured reliably.

Comparison between equivalent widths measured from the two plates shows a mean difference of 0($\pm3$)\,\AA \/ for the 11 galaxies for which equivalent widths were measured on both plates. The mean measured equivalent widths ($W_{\lambda}$) are listed in column 3 of Table \ref{ewtab}. A colon appended to the equivalent width value indicates some additional uncertainty in fitting the continuum.

H$\alpha$+[\ion{N}{ii}] fluxes were measured in arbitrary units of photographic density, which is approximately proportional to intensity.
We then used the flux calibration of \citeN{Bes} to give H$\alpha$+[\ion{N}{ii}] fluxes from the $R$ magnitudes, where available in the literature, and the equivalent width measurements. These flux values were used to calibrate the photographic fluxes, which were then converted to luminosities using corrections for galactic extinction from \citeN{Burstein} and the NASA Extragalactic Database (NED), the internal extinction correction given by \citeN{rc3}, and a distance  corresponding to the cluster mean redshift for cluster members, or individual galaxy redshifts for background galaxies, using the velocity correction for Virgocentric infall given in R89 \cite{r87b}. $H_0$ was taken as 50 km s$^{-1}$ Mpc$^{-1}$. The scatter in this calibration is $\sim0.08$ in the log, corresponding to an uncertainty of $\sim$20\% in luminosity.

The \mbox{H$\alpha$+[\ion{N}{ii}]} luminosities from this calibration were corrected to give H$\alpha$ luminosities using the conversion of \mbox{$W_{\lambda}$(H$\alpha$+[\ion{N}{ii}])=1.33$W_{\lambda}$(H$\alpha$)} from \citeN{kenn83}. This was then converted to an effective star formation rate using the conversion factor of \citeN{alhe} of \mbox{$L$(H$\alpha$)/SFR=3.10$\times$10$^{41}$erg  s$^{-1}$/M$_{\odot}$  yr$^{-1}$}, which corresponds to a \citeN{salpeter} IMF with upper and lower mass cutoffs of \mbox{125 M$_{\odot}$} and \mbox{0.1 M$_{\odot}$} respectively. This conversion is highly IMF dependent.

\begin{table}
\centering
\caption{\label{ewtab} Global H$\alpha$+[\ion{N}{ii}] equivalent widths and fluxes}
\begin{tabular}{rrlcl}
\\
\hline
\\
No. &Galaxy name &$W_{\lambda}$&$f\times 10^{14}$&SFR\\
 & & \AA & ergs cm$^{-2}$ s$^{-1}$&M$_{\odot}$ yr$^{-1}$\\
\\
(1)&(2)&(3)&(4)&(5)\\
\\
\hline
\\
1&R89 57 	&19:\hspace{-0.7ex}            	& \hspace{1.0ex}6 &    0.1\\      
5&ESO 501-G17 	&23:\hspace{-0.7ex}	    	& 16 &     0.4\\  
6&ESO 436-IG42 	&96			    	& 85 &     1.4\\  
7&ESO 436-G43 	&52			    	& 16 &     1.8\\  
8&ESO 437-IG3 	&59			    	& 17 &     0.4\\  
9&ESO 501-G32 	&42			    	& 40 &     0.6\\  
10&R89 281 	&49			    	& 19 &     0.3\\  
12&ESO 501-G45 	&21:\hspace{-0.7ex}	    	& 10 &     0.2\\  
16&ESO 501-IG61 &31			    	& 11 &     0.2\\  
18&ESO 501-G65 	&53			    	& 92 &     1.7\\  
22&ESO 501-G70 	&19:\hspace{-0.7ex}	    	& \hspace{1.0ex}8 &     1.3\\  
31&IRAS 10448-2839 	&38:\hspace{-0.7ex} 	& 37 &     4.8\\  
32&NGC 3393 	& 39			    	& \hspace{-1.0ex}148 &    \dots\\
\\
\hline
\\
\end{tabular}
\end{table}

Calibrated \mbox{H$\alpha$+[\ion{N}{ii}]} fluxes and effective star formation rates are listed in columns 4 and 5 of Table \ref{ewtab}. A star formation rate is not listed for NGC 3393 as it is a Seyfert. Despite the uncertainties in the star formation rate determination, the derived values of between a tenth of a solar mass and a few solar masses per year lie within the range expected for normal spirals. Nevertheless, a few galaxies appear to have an anomalously high star formation rate for their morphological type (see \S \ref{frac}).

\subsection{Cluster emission-line galaxies}
\label{frac}
\noindent

From the distribution of detected emission-line galaxies with magnitude
it appears that the complete survey sample (V $\le$ 16.65, r $<$
2\degr) extends somewhat below the detection limit of the H$\alpha$
survey, although this detection limit is not well defined.  For the
magnitude ranges, 12 $<$ V $\le$ 14, 14 $<$ V $\le$ 16, and V $>$
16 the percentages of galaxies, types Sa and later, detected in emission
are 24\%, 23\% and 6\% respectively. We estimate that emission-line galaxy
detections are reasonably complete to V=16.  Accordingly in what follows,
we restrict discussion to a subsample of the complete sample, with V $\le$
16, r $<$ 2\degr.  There are some 180 galaxies in this subsample which
have been surveyed for H$\alpha$ emission, of which
35 are classified in R89 as E or E/S0, 62 as S0 or S0/a, and 65 as Sa or later, 
with 8 being unclassified.  The
percentages of the three type groups detected in emission are 6\%, 4\% and 
23\% respectively.  As expected, only a small percentage of early type
galaxies were detected in emission.

\begin{figure}
\psfig{figure=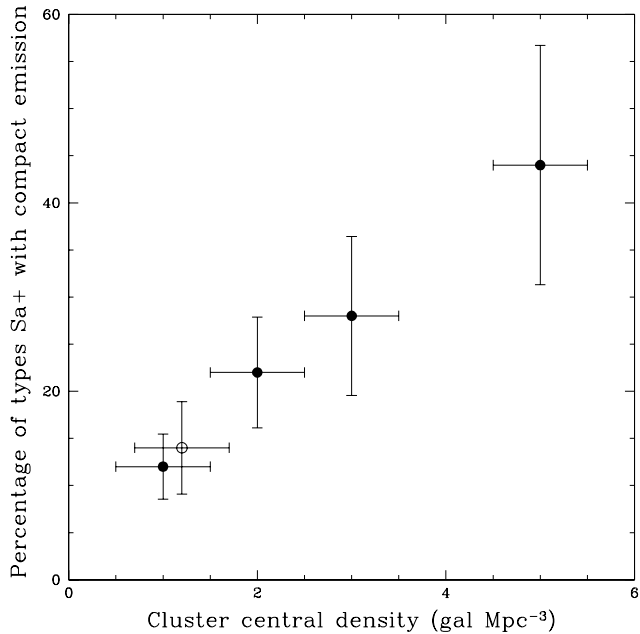,width=0.5\textwidth,height=0.5\textwidth}
\caption{\label{fracfig} The percentage of galaxies of types Sa and later showing compact emission plotted as a function of cluster central galaxy density. Filled circles indicate the eight clusters studied by Moss \& Whittle (1997), the open circle indicates Abell 1060. Error bars indicate Poisson errors.}
\end{figure}

Previous work \cite{moss93} has shown the usefulness of a distinction between {\em diffuse} emission
described by the concentration classes D (diffuse) and VD (very diffuse), and {\em compact} emission 
described by the other three classes VC (very concentrated), C (concentrated), and N (normal). It has 
been found that compact emission is strongly associated with a disturbed morphology of the galaxy, and 
most likely results from tidally-induced star formation from galaxy--galaxy or cluster--galaxy 
interactions. Furthermore there is a strong
correlation between cluster 
mean central galaxy density and the fraction of galaxies of types Sa and later
with compact emission \cite{moss96}.  
This is illustrated in Fig. \ref{fracfig}, 
where the central galaxy density is 
calculated from the number of galaxies with absolute magnitude $M_{T,0}\leq-20.4$ 
within $0.5r_{A}$ of the cluster centre, corrected for the effects of foreground 
and background contamination and cluster galaxies projected onto the central region. 
For Abell 1060, the central galaxy density calculated in this way is approximately 
1.2 Mpc$^{-3}$. However, the actual value is likely to be lower than this as the 
correction for projection was less accurate than was possible for more distant clusters. 
Within 120 arcmin ($\sim 1\,r_{A}$) of the cluster centre 8 galaxies (14\%) of types Sa 
and later have compact emission. This is in good agreement with the fraction of 12\% 
for the lowest density bin in Fig. \ref{fracfig}, indicating 
that the number of spirals detected with compact emission in Abell 1060 agrees
with the expected value for a cluster of its richness.

A detailed comparison between the star formation rates of cluster galaxies
in Abell 1060 and corresponding rates for field galaxies is not possible
due both to the small
sample of detected emission-line galaxies and incomplete measurements of
H$\alpha$ equivalent widths.  Nevertheless it may be noted that a 
significant fraction of the detected cluster emission-line galaxies are
early-type spirals which may be surprising in view of the detection limit
of 
the survey technique of $\sim$ 20 \AA\, (see MWI), and the expectation that
H$\alpha$
equivalent widths for galaxies of types Sab and earlier in the field are less than 20
\AA\, \cite{kennkent}.
Indeed, the three galaxies ESO 501-G17, ESO 501-G45, and R89 281, 
typed Sa or earlier by R89 and \citeN{wang}, and known cluster members,
have measured equivalent widths 
greater than 20\,\AA, and can therefore be considered to have unusually high star 
formation rates.  The cluster elliptical ESO 436-IG42 also has a very high 
equivalent width of 96\,\AA \,(see \S \ref{indiv}).
These results are consistent with an enhanced star formation
rate in early-type cluster spirals found previously for other clusters (see \citeNP{moss93}).

\subsection{Notes on individual galaxies}
\label{indiv}

{\em No. 6, ESO 436-IG42: } This galaxy is clearly interacting with a nearby companion; they are joined by a bridge of material \cite{lau}. The fact that it is classified as an elliptical and exhibits compact emission points to the possibility of a centrally-concentrated burst of star formation triggered by this interaction.

{\em No. 13, NGC 3314A: }This is the foreground member of the remarkable superposed galaxy pair NGC 3314, discussed in detail by \citeN{n3314} and \citeN{r82}, which consists of two spiral galaxies of comparable angular size, one face-on and the other more nearly edge-on, whose centres are superposed almost exactly along the line of sight. Two redshifts have been determined for the pair from optical and 21 cm line measurements of 2\,851 and \mbox{4\,641\,km\,s$^{-1}$}. Both sets of authors identify NGC 3314A as both the foreground object (dust lanes in its disk obscure NGC 3314B) and as having the lower recession velocity. The two-plate redshift estimate (see \S \ref{twoplate}) of \mbox{2\,600\,$\pm$\,340\,km\,s$^{-1}$} confirms that emission is detected in the foreground NGC 3314A, in agreement with the spectrum of \citeN{n3314}.

{\em Nos. 16 \& 17, ESO 501-IG61 \& 501-G62:} These two galaxies are seen in close separation on the sky, \mbox{$\sim$ 1 arcmin} apart, and might consequently be taken for an interacting pair. As they lie outside the overlap area, a two-plate redshift measurement was not possible. However, a less accurate estimate can be made using a single plate, giving redshifts of \mbox{$\sim$ 3\,200 km\,s$^{-1}$} for ESO 501-IG61 and \mbox{$\sim$ 10\,200 km\,s$^{-1}$} for ESO 501-G62, with an adopted uncertainty of \mbox{$\sim$ 550 km\,s$^{-1}$} (MWI). This is accurate enough to identify ESO 501-IG61 as a possible cluster member and ESO 501-G62 as a non-member, and to confirm that they are not an interacting pair as might otherwise be inferred from their close separation. The R89 value of \mbox{3\,597 km\,s$^{-1}$} has been adopted for the redshift of ESO 501-IG61.

\section{\label{discus}Discussion and conclusions}

\noindent We have surveyed the Hydra cluster, Abell 1060, for star-forming 
galaxies using an objective prism technique to detect H$\alpha$ emission. 
We detect a total of 33 ELGs in the survey area, all of which have been 
identified with previously-known objects from one or more of a variety of 
catalogues. Radial velocities have been determined for 7 ELGs
without previous determinations, and measurements have been made of global 
H$\alpha$+N[II] equivalent width and flux values for 13 ELGs.  For a 
complete galaxy sample (n=180) with $V \le 16$ within 2$^{\circ}$ of the 
cluster centre, 24 galaxies are detected in emission, of which 18 are 
cluster members.

In accord with previous work (MWI) we have classified the appearance of the 
emission as compact or diffuse.  It has previously been found that compact
emission is associated with a disturbed galaxy morphology, and is most likely
the result of tidally-induced star formation either by galaxy--galaxy or 
cluster--galaxy interactions (see \citeNP{moss93}). Furthermore the fraction of
spirals in a cluster with compact emission has previously been found to correlate
with cluster richness (see \citeNP{moss96}).  Using the complete
galaxy sample we show that the fraction of spirals detected with compact
emission in Abell 1060 is consistent with this correlation.

Finally, although the small sample of detected cluster ELGs in Abell 1060 precludes a 
detailed comparison of star formation rates between these cluster galaxies and
corresponding types in the field, it is to be noted that at least three early-type cluster spirals
(types S0 to Sa) and an elliptical have global H$\alpha$ equivalent widths greater than 20 \AA, which would be
anomalously high for these galaxy types in the field.  This is consistent with an 
enhanced star formation rate in early-type cluster spirals found in previous work \cite{moss93}.

\begin{acknowledgements}

We thank V.M. Blanco who generously provided the
plate material on which this study is based.
SMB is supported by the Particle Physics and Astronomy Research Council.
CM thanks the Institute of Astronomy, Cambridge for support during the
course of this project. 
This research has made use of the NASA/IPAC Extragalactic Database (NED) 
which is operated by the Jet Propulsion Laboratory, Caltech, 
under contract with the National Aeronautics and Space Administration.
Cerro Tololo Inter-American Observatory, National Optical Astronomy 
Observatories, are operated by the Association of Universities for 
Research in Astronomy, under contract with the National Science Foundation.

\end{acknowledgements}

\end{document}